\documentclass{article}%
\usepackage{amsmath}
\usepackage{amsfonts}
\usepackage{amssymb}
\usepackage{graphicx}%
\providecommand{\U}[1]{\protect\rule{.1in}{.1in}}
\begin{document}

\title{Elicitation, measuring bias, checking for prior-data conflict and inference with
a Dirichlet prior}
\author{Michael Evans, Irwin Guttman and Peiying Li\\Department of Statistical Sciences\\University of Toronto}
\date{}
\maketitle

\noindent\textit{Abstract}: Methods are developed for eliciting a Dirichlet
prior based upon bounds on the individual probabilities that hold with virtual
certainty. This approach to selecting a prior is applied to a contingency
table problem where it is demonstrated how to assess the bias in the prior as
well as how to check for prior-data conflict. It is shown that the assessment
of a hypothesis via relative belief can easily take into account what it means
for the falsity of the hypothesis to correspond to a difference of practical
importance and provide evidence in favor of a hypothesis. \medskip


\section{Introduction}

Perhaps the most basic statistical model is the multinomial$(n,p_{1}%
,\ldots,p_{k})$ where $n\in%
%TCIMACRO{\U{2115} }%
%BeginExpansion
\mathbb{N}
%EndExpansion
,(p_{1},\ldots,p_{k})\in S_{k}=\{(x_{1},\ldots,x_{k}):x_{i}\geq0$ and
$x_{1}+\cdots+x_{k}=1\},S_{k}$ is the $(k-1)$-dimensional simplex and
$(p_{1},\ldots,p_{k})$ is unknown. This arises from an i.i.d. sample from the
multinomial$(1,p_{1},\ldots,p_{k})$ distribution. The goal is then inference
about the unknown value of $(p_{1},\ldots,p_{k}).$

Bayesian inference requires a prior and the Dirichlet$(\alpha_{1}%
,\ldots,\alpha_{k}),$ for some choice of hyperparameters $\alpha_{1}%
,\ldots,\alpha_{k},$\ is a convenient choice due to its conjugacy. To employ
such a prior it is necessary to have an easy to use elicitation algorithm. The
purpose of this paper is to develop such an algorithm, to show how the chosen
prior can be assessed with respect to the bias that it induces, to check
whether the prior conflicts with the data, to show how to modify the prior
when such a conflict is encountered and to implement inferences using the
prior based on a measure of statistical evidence.

In Section 2 an elicitation algorithm is developed for the Dirichlet. In
Section 3 the bias in the prior is discussed and in Section 4 the issue of
prior-data conflict and possible modification of the prior is addressed.
Section 5 deals with inference for the multinomial based on the relative
belief ratio as a measure of evidence. This presents a full treatment of a
statistical analysis for the multinomial although it is assumed that the
multinomial model is correct. Strictly speaking, provided the data is
available, it should also be checked that the initial sample is i.i.d. from a
multinomial$(1,p_{1},\ldots,p_{k})$ distribution, perhaps using a multivariate
version of a runs test, but this is not addressed here.

Throughout the paper the following example, taken from Snedecor and Cochran
(1967), is considered as a practical application of the methodology.\smallskip

\noindent\textbf{Example 1}. \textit{Assessing independence}

\noindent Individuals were classified according to their blood type $Y$
($O,A,B,$ and $AB$,\ although the $AB$ individuals were eliminated, as they
were small in number) and also classified according to $X,$ their disease
status (peptic ulcer = $P$, gastric cancer = $G$, or control = $C$). So there
are three populations; namely, those suffering from a peptic ulcer, those
suffering from gastric cancer, and those suffering from neither and it is
assumed that the individuals involved in the study can be considered as random
samples from the respective populations. The data are in Table 1 and the goal
is to determine whether or not $X$ and $Y$ are independent. So the counts are
assumed to be multinomial$(8766,p_{11},p_{12},p_{13},p_{21},p_{22}%
,p_{23},p_{31},p_{32},p_{33})$ where the first index refers to $X$ and the
second to $Y\,$\ and with a relabelling of the categories, e.g. $X=G$ is
relabeled as $X=2.$%

%TCIMACRO{\TeXButton{B}{\begin{table}[tbp] \centering}}%
%BeginExpansion
\begin{table}[tbp] \centering
%EndExpansion
$%
\begin{tabular}
[c]{|c|rrrr|}\hline
& $Y=O$ & $Y=A$ & $Y=B$ & Total\\\hline
$X=P$ & $983$ & $679$ & $134$ & $1796$\\
$X=G$ & $383$ & $416$ & $84$ & $883$\\
$X=C$ & $2892$ & $2625$ & $570$ & $6087$\\
Total & $4258$ & $3720$ & $788$ & $8766$\\\hline
\end{tabular}
\ $\caption{The data in Example 1.}\label{TableKey}%
%TCIMACRO{\TeXButton{E}{\end{table}}}%
%BeginExpansion
\end{table}%
%EndExpansion

Using the chi-squared test, the null hypothesis of no relationship is rejected
with a value of the chi-squared\ statistic of $40.54$\ and a $p$-value of
$0.0000$. Table 2 gives the estimated cell probabilities based on the full
multinomial as well as the estimated cell probabilities based on independence
between the $X$ and $Y.$ The difference between the two tables is very small
and of questionable practical significance. For example, the largest
difference between corresponding cells is $0.012$ and, as a natural measure of
difference between two distributions, the estimated Kullback-Leibler
divergence, based on the raw data, is estimated as $0.002.$ This suggests that
in reality the deviation from independence is not meaningful. The cure for
this is that, in assessing any hypothesis, it is necessary to say what size of
deviation $\delta$ from the null is of practical significance and take this
into account when performing the test. This arises as a natural aspect of the
relative belief approach to this problem and will be discussed in Section 3,
where a very different conclusion is reached in this example.%

%TCIMACRO{\TeXButton{B}{\begin{table}[tbp] \centering}}%
%BeginExpansion
\begin{table}[tbp] \centering
%EndExpansion
$%
\begin{tabular}
[c]{|cccc|c|ccc|}\hline
\multicolumn{1}{|c|}{Full} & $Y=O$ & $Y=A$ & $Y=B$ & Ind. & $Y=O$ & $Y=A$ &
$Y=B$\\\hline
\multicolumn{1}{|c|}{$X=P$} & \multicolumn{1}{r}{$0.112$} &
\multicolumn{1}{r}{$0.077$} & \multicolumn{1}{r|}{$0.015$} & $X=P$ & $0.100$ &
$0.087$ & $0.018$\\
\multicolumn{1}{|c|}{$X=G$} & \multicolumn{1}{r}{$0.043$} &
\multicolumn{1}{r}{$0.047$} & \multicolumn{1}{r|}{$0.009$} & $X=G$ & $0.049$ &
$0.043$ & $0.009$\\
\multicolumn{1}{|c|}{$X=C$} & \multicolumn{1}{r}{$0.330$} &
\multicolumn{1}{r}{$0.299$} & \multicolumn{1}{r|}{$0.065$} & $X=C$ & $0.337$ &
$0.295$ & $0.062$\\\hline
\end{tabular}
$%
\caption{The  estimated cell probabilities in Example 1 based on the full and independence models.}\label{TableKey copy(1)}%
%TCIMACRO{\TeXButton{E}{\end{table}}}%
%BeginExpansion
\end{table}%
%EndExpansion

\section{Elicitation}

A key component of a Bayesian statistical analysis is the choice of the prior.
For this it is recommended that an elicitation algorithm be used so that the
selection of the prior be based upon what is known about problem under study.
Typically this will involve some knowledge of what kind of values are expected
for the data as these arise via some measurement process. In the context of
the Dirichlet this knowledge will take the form of how likely a success is
expected on each of the $k$ categories being counted. Of course, there can be
a variety of elicitation algorithms that are appropriate. Our approach here is
to develop one that is simple to use and results in an appropriate expression
of belief. Discussions about the process of elicitation for general problems
can be found in Gathwaite at al. (2005) and O'Hagan et al. (2006).

Consider first the situation where $k=2$ and the prior $\Pi_{\alpha_{1}%
,\alpha_{2}}$ on $p_{1}$ is beta$(\alpha_{1},\alpha_{2})$. Suppose it is known
with `virtual certainty' that $l_{1}\leq p_{1}\leq u_{1}$ where $l_{1}%
,u_{1}\in\lbrack0,1]$ are known. This immediately implies that $1-u_{1}\leq
p_{2}=1-p_{1}\leq1-l_{1}$ with virtual certainty. Here `virtual certainty' is
interpreted to mean that the true value of $p_{1}$ is in the interval
$[l_{1},u_{1}]$ with high prior probability $\gamma$, say $\gamma=0.99.$ So
this restricts the prior to those values of $(\alpha_{1},\alpha_{2})$
satisfying $\Pi_{\alpha_{1},\alpha_{2}}([l_{1},u_{1}])=\gamma.$ To completely
determine $(\alpha_{1},\alpha_{2})$ another condition is added, namely, it is
required that the mode of the prior be at the point $\xi\in\lbrack l_{1}%
,u_{1}]$ as this allows the placement of the primary amount of the prior mass
at an appropriate place within $[l_{1},u_{1}].$ For example, a natural choice
of the mode in this context is $\xi=(l_{1}+u_{1})/2$, namely, the midpoint of
the interval. When $\alpha_{1},\alpha_{2}\geq1$ the mode of the beta$(\alpha
_{1},\alpha_{2})$ occurs at $\xi=(\alpha_{1}-1)/\tau$ where $\tau=\alpha
_{1}+\alpha_{2}-2.$ There is thus a 1-1 correspondence between the values
$(\alpha_{1},\alpha_{2})$ and $(\xi,\tau)$ given by $\alpha_{1}=1+\tau
\xi,\alpha_{2}=1+\tau(1-\xi).$ Therefore, after specifying the mode, only the
scaling of the beta prior is required through the choice of $\tau.$ The value
$\tau$ is completely determined by $\Pi_{\alpha_{1},\alpha_{2}}([l_{1}%
,u_{1}])=\gamma$ provided that $u_{1}-l_{1}\leq\gamma$ as it is easy to see
that $\Pi_{1+\tau\xi,1+\tau(1-\xi)}([l_{1},u_{1}])\uparrow1$ as $\tau
\uparrow\infty.$ Note that the restriction $\alpha_{1},\alpha_{2}\geq1$ is
natural as this avoids singularities at 0 or 1. If $u_{1}-l_{1}>\gamma,$ then
the requirement can be relaxed to requiring $(\alpha_{1},\alpha_{2})$ satisfy
$\Pi_{\alpha_{1},\alpha_{2}}([l_{1},u_{1}])\geq\gamma,$ so the beta$(1,1)$
suffices or a larger value of $\gamma$ can be chosen.

Supposing $u_{1}-l_{1}\leq\gamma,$ it is then straightforward to solve for
$\tau$ via an iterative algorithm. To start set $\tau_{0}=0,$ which implies
$(\alpha_{1},\alpha_{2})=(1,1)\ $\ and $\Pi_{1+\tau_{0}\xi,1+\tau_{0}(1-\xi
)}([l_{1},u_{1}])=u_{1}-l_{1},$ find $\tau_{1}$ such that $\Pi_{1+\tau_{1}%
\xi,1+\tau_{1}(1-\xi)}([l_{1},u_{1}])>\gamma$ and then proceed iteratively via
the bisection root finding algorithm.\smallskip

\noindent\textbf{Example 2}. \textit{Determining a beta prior.}

Suppose that $[l_{1},u_{1}]=(0.25,0.75),\xi=0.5$ and $\gamma=0.99.$ The
solution obtained via the iterative algorithm is then $\tau=22.0$ where the
iteration is stopped when $|\Pi_{1+\tau_{i}\xi,1+\tau_{i}(1-\xi)}([l_{1}%
,u_{1}])-\gamma|\leq0.005.$ This took 7 iterations and the prior is given by
$(\alpha_{1},\alpha_{2})=(12.0,12.0)$ and $[l_{1},u_{1}]$ contains $0.993$ of
the prior probability. If instead of $0.005$ the error tolerance for stopping
was set equal to $0.001$, then the solution $\tau=22.04$ and $(\alpha
_{1},\alpha_{2})=(12.02,12.02)$ was obtained after 20 iterations with
$[l_{1},u_{1}]$ containing $0.990$ of the prior probability.\smallskip

The approach to eliciting a beta prior seems very natural and allows for a
great deal of flexibility in where the prior allocates the bulk of its mass in
$[0,1].$ The question, however, is how to generalize this to the
Dirichlet$(\alpha_{1},\ldots,\alpha_{k})$ prior. As will be seen, it is
necessary to be careful about how $(\alpha_{1},\ldots,\alpha_{k})$ is
elicited. Again we make the restriction that each $\alpha_{i}\geq1$ to avoid
singularities for the prior on the boundary.

It seems quite natural to think about putting probabilistic bounds on the
$p_{i}$ such as requiring $l_{i}\leq p_{i}\leq u_{i}$ with high probability,
for fixed constants $l_{i},u_{i},$ to reflect what is known with `virtual
certainty' about $p_{i}.$ For example, it may be known that $p_{i}$ is very
small and so we put $l_{i}=0$, choose $u_{i}$ small and require that
$p_{i}\leq u_{i}$ with prior probability at least $\gamma.$ While placing
bounds like this on the $p_{i}$ seems reasonable, such an approach can result
in a complicated shape for the region that is to contain the true value of
$(p_{1},\ldots,p_{k})$ with virtual certainty. This complexity can make the
computations associated with inference very difficult. In fact it can be hard
to determine exactly what the full region is. As such, it seems better to use
an elicitation method that fits well with the geometry of the Dirichlet
family. If it is felt that more is known a priori than an Dirichlet prior can
express, then it is appropriate to contemplate using some other family of
priors. Given the conjugacy property of Dirichlet priors, which vastly
simplifies many computations, the focus here is on devising elicitation
algorithms that work well with this family. First we consider elicitation
approaches for this problem that have been presented in the literature.

Chaloner and Duncan (1987) discuss an iterative elicitation algorithm based on
specifying characteristics of the prior predictive distribution of the data
which is Dirichlet-multinomial. Regazzini and Sazonov (1999) discuss an
elicitation algorithm which entails partitioning the simplex, prescribing
prior probabilities for each element of the partition and then selecting a
mixture of Dirichlet distributions as the prior such that this prior has
Prohorov distance less than some $\epsilon>0$ from the true prior associated
with de Finetti's representation theorem. Both of these approaches are
complicated to implement. Closest to the method presented here is that
discussed in Dorp and Mazzuchi (2003) where $(\alpha_{1},\ldots,\alpha_{k})$
is specified by choosing $i\in\{1,\ldots,k\},$ stating two prior quantiles
$(p_{\gamma_{i1}},p_{\gamma_{i2}})$ where $0<\gamma_{i1}<\gamma_{i2}<1$ for
$p_{i}$ and specifying prior quantile $p_{\gamma_{j}}$ for $p_{j}$ for each
$j\neq i,k.$ So there are $k$ constraints that the Dirichlet$(\alpha
_{1},\ldots,\alpha_{k})$ has to satisfy and an algorithm is provided for
computing $(\alpha_{1},\ldots,\alpha_{k}).$ Drawbacks include the fact that
the $p_{i}$ are not treated symmetrically as there is a need to place two
constraints on one of the probabilities, $p_{k}$ is treated quite differently
than the other probabilities, precise quantiles need to be specified and
values $\alpha_{i}<1$ can be obtained which induce singularities in the prior.
Furthermore, it is not at all clear what these constraints say about the joint
prior on $(p_{1},\ldots,p_{k})$ as this elicitation does not take into account
the dependencies that occur necessarily among the $p_{i}.$

A simpler approach to elicitation is now developed. There are several versions
depending on whether lower or upper bounds are placed on the $p_{i}.$ We start
with the situation where a lower bound is given for each $p_{i}$ as this
provides the basic idea for the others. Generally the elicitation process
allows for a single lower or upper bound to be specified for each $p_{i}.$
These bounds specify a subsimplex of the simplex $S_{k}$ with all edges of the
same length. As will be seen, this implicitly takes into account the
dependencies among the $p_{i}.$ With such a region determined, it is
straightforward to determine $(\alpha_{1},\ldots,\alpha_{k})$ such that the
subsimplex contains $\gamma$ of the prior probability for $(p_{1},\ldots
,p_{k}).$

Note that a $(k-1)$-dimensional simplex can be specified by specifying $k$
distinct points in $R^{k},$ say $\mathbf{a}_{1},\ldots,\mathbf{a}_{k},$ and
then taking all convex combinations of these points. This simplex will be
denoted as $S(\mathbf{a}_{1},\ldots,\mathbf{a}_{k})=\{\sum_{i=1}^{k}%
c_{i}\mathbf{a}_{i}:c_{i}\geq0$ with $c_{1}+\cdots+c_{k}=1\}.$ So
$S_{k}=S(\mathbf{e}_{1},\ldots,\mathbf{e}_{k})$ and it is clear that
$S(\mathbf{a}_{1},\ldots,\mathbf{a}_{k})\subset S(\mathbf{e}_{1}%
,\ldots,\mathbf{e}_{k})$ whenever $\mathbf{a}_{1},\ldots,\mathbf{a}_{k}\in
S(\mathbf{e}_{1},\ldots,\mathbf{e}_{k}).$ The centroid of $S(\mathbf{a}%
_{1},\ldots,\mathbf{a}_{k})$ is equal to $CS(\mathbf{a}_{1},\ldots
,\mathbf{a}_{k})=\sum_{i=1}^{k}\mathbf{a}_{i}/k.$

\subsection{Lower bounds on the probabilities}

For this we ask for a set of lower bounds $l_{1},\ldots,l_{k}\in\lbrack0,1]$
such that $l_{i}\leq p_{i}$ for $i=1,\ldots,k.$ To make sense there is only
one additional constraint that the $l_{i}$ must satisfy, namely,
$L_{1:k}=l_{1}+\cdots+l_{k}\leq1.$ If $L_{1:k}=1,$ then it is immediate that
$p_{i}=l_{i},$ otherwise $p_{1}+\cdots+p_{k}>1.$ So the $p_{i}$ are completely
determined when $L_{1:k}=1.$ Attention is thus restricted to the case where
$L_{1:k}<1.$ The following result then holds.\smallskip

\noindent\textbf{Theorem 1.} Specifying the lower bounds $l_{1},\ldots
,l_{k}\in\lbrack0,1]$ such that $l_{i}\leq p_{i}$ for $i=1,\ldots,k$ and
\begin{equation}
L_{1:k}<1, \label{ineq1}%
\end{equation}
prescribes $S(\mathbf{a}_{1},\ldots,\mathbf{a}_{k})\subset S_{k}$ where
$\mathbf{a}_{i}=(l_{1},\ldots,l_{i-1},u_{i},l_{i+1},\ldots,l_{k})$ and
\begin{equation}
u_{i}=1-\sum_{j\neq i}l_{j}. \label{upperbd}%
\end{equation}
The edges of $S(\mathbf{a}_{1},\ldots,\mathbf{a}_{k})$ each have length
$\sqrt{2}(1-L_{1:k})\ $and $S(\mathbf{a}_{1},\ldots,\mathbf{a}_{k}%
)=\{(p_{1},\ldots,p_{k}):p_{1}+\cdots+p_{k}=1,l_{i}\leq p_{i}\leq
u_{i},i=1,\ldots,k\}.$

\noindent Proof: Note that (\ref{ineq1}) implies that $p_{i}=1-\sum_{j\neq
i}p_{j}\leq1-\sum_{j\neq i}l_{j}=u_{i},$ and so stating the lower bounds
implies a set of upper bounds, and also $l_{i}<u_{i}\leq1.$ Consider now the
set $S=\{(p_{1},\ldots,p_{k}):p_{1}+\cdots+p_{k}=1,l_{i}\leq p_{i}\leq
u_{i},i=1,\ldots,k\}$ and note that $\mathbf{a}_{i}\in S$ for $i=1,\ldots,k.$
For $c_{i}\geq0$ with $c_{1}+\cdots+c_{k}=1,$ then $(p_{1},\ldots,p_{k}%
)=\sum_{i=1}^{k}c_{i}\mathbf{a}_{i}\in S$ since, for example, the first
coordinate satisfies $p_{1}=c_{1}u_{1}+(\sum_{i=2}^{k}c_{i})l_{1}=c_{1}%
u_{1}+(1-c_{1})l_{1}$ so $l_{1}\leq p_{1}\leq u_{1}.$ Therefore $S(\mathbf{a}%
_{1},\ldots,\mathbf{a}_{k})\subset S.$

If $(p_{1},\ldots,p_{k})\in S,$ then $p_{i}=c_{i}^{\ast}l_{i}+(1-c_{i}^{\ast
})u_{i}$ where $c_{i}^{\ast}\in\lbrack0,1].$ Now $1=p_{1}+\cdots+p_{k}%
=\sum_{i=1}^{k}c_{i}^{\ast}l_{i}+\sum_{i=1}^{k}(1-c_{i}^{\ast})u_{i}%
=\sum_{i=1}^{k}c_{i}^{\ast}l_{i}+\sum_{i=1}^{k}(1-c_{i}^{\ast})\left(
l_{i}+1-L_{1:k}\right)  =L_{1:k}+\{\sum_{i=1}^{k}(1-c_{i}^{\ast})\}\left(
1-L_{1:k}\right)  $ and so $\sum_{i=1}^{k}(1-c_{i}^{\ast})=1.$ For
$(p_{1},\ldots,p_{k})=\sum_{j=1}^{k}(1-c_{j}^{\ast})\mathbf{a}_{j}$ we have
$p_{i}=(\sum_{j\neq i}(1-c_{j}^{\ast}))l_{i}+(1-c_{i}^{\ast})u_{i}=c_{i}%
^{\ast}l_{i}+(1-c_{i}^{\ast})u_{i}.$ This proves that $S\subset S(\mathbf{a}%
_{1},\ldots,\mathbf{a}_{k})$ and so we have $S(\mathbf{a}_{1},\ldots
,\mathbf{a}_{k})=S.$

Finally note that $||\mathbf{a}_{i}-\mathbf{a}_{j}||^{2}=(u_{i}-l_{i}%
)^{2}+(u_{i}-l_{j})^{2}=2(1-L_{1:k})^{2}$ and so $S(\mathbf{a}_{1}%
,\ldots,\mathbf{a}_{k})$ has edges all of the same length. This completes the proof.

\subsection{Upper bounds on the probabilities}

Of course, it may be that prior beliefs are instead expressed via upper bounds
on the probabilities or a mixture of upper and lower bounds. The case of all
upper bounds is considered first. Our goal is to specify the upper bounds in
such a way that these lead unambiguously to lower bounds $l_{1},\ldots
,l_{k}\in\lbrack0,1]$ satisfying (\ref{ineq1}) and so to the simplex
$S(\mathbf{a}_{1},\ldots,\mathbf{a}_{k}).$

Suppose then that we have the upper bounds $u_{1},\ldots,u_{k}\in\lbrack0,1]$
such that $p_{i}\leq u_{i}.$ It is clear then that $l_{1},\ldots,l_{k}$ must
satisfy the system of linear equations given by (\ref{upperbd}) as well as
$0\leq l_{i}\leq u_{i}$ for $i=1,\ldots,k$ and (\ref{ineq1}). So the $l_{i}$
must satisfy
\begin{equation}
\mathbf{u=1}_{k}-\left(
\begin{array}
[c]{cccc}%
0 & 1 & \ldots & 1\\
1 & 0 & \ldots & 1\\
\vdots & \vdots & \vdots & \vdots\\
1 & 1 & \ldots & 0
\end{array}
\right)  \boldsymbol{l\,}\mathbf{=1}_{k}+(I_{k}-\mathbf{1}_{k}\mathbf{1}%
_{k}^{\prime})\boldsymbol{l} \label{mateq1}%
\end{equation}
where $\mathbf{1}_{k}$ is the $k$-dimensional vector of 1's and $I_{k}$ is the
$k\times k$ identity. Noting that $(I_{k}-\mathbf{1}_{k}\mathbf{1}_{k}%
^{\prime})^{-1}=I_{k}-(k-1)^{-1}\mathbf{1}_{k}\mathbf{1}_{k}^{\prime},$ it is
immediate that
\begin{equation}
\boldsymbol{l\,}\mathbf{=(}I_{k}-(k-1)^{-1}\mathbf{1}_{k}\mathbf{1}%
_{k}^{\prime})(\mathbf{u}-\mathbf{1}_{k}\mathbf{).} \label{eq1}%
\end{equation}
Note that this requires that $k\geq2$ as is always the case.

Putting $U_{1:k}=\sum_{j=1}^{k}u_{j},$ then (\ref{eq1}) implies $L_{1:k}%
=(k-U_{1:k})/(k-1)$ and so $0\leq L_{1:k}<1$ provided $U_{1:k}$ satisfies
\begin{equation}
1<U_{1:k}\leq k. \label{ineq2}%
\end{equation}
From (\ref{eq1})
\begin{equation}
l_{i}=(u_{i}-1)-\frac{U_{1:k}-k}{k-1}=u_{i}+\frac{1-U_{1:k}}{k-1} \label{eq2}%
\end{equation}
and, for $i=1,\ldots,k,$ this implies that $l_{i}\geq0$ iff
\begin{equation}
u_{i}\geq\frac{U_{1:k}-1}{k-1}. \label{ineq3}%
\end{equation}
Also, when (\ref{ineq2}) is satisfied, then $l_{i}<u_{i}$ for $i=1,\ldots,k.$
This completes the proof of the following result.\smallskip

\noindent\textbf{Theorem 2.} Specifying upper bounds $u_{1},\ldots,u_{k}%
\in\lbrack0,1],$ such that $p_{i}\leq u_{i}$ for $i=1,\ldots,k,$ satisfying
inequalities (\ref{ineq2}) and (\ref{ineq3}), determines \ the lower bounds
$l_{1},\ldots,l_{k},$ given by (\ref{eq2}), which determine the simplex
$S(\mathbf{a}_{1},\ldots,\mathbf{a}_{k})$ defined in Theorem 1.\smallskip

The difficult aspect of this approach to elicitation is to make sure the upper
bounds satisfy (\ref{ineq2}) and (\ref{ineq3}). If we take $u_{1}=\cdots
=u_{k}=u\geq1/k,$ then (\ref{ineq2}) is satisfied and $(k-1)u\geq ku-1$
implies that (\ref{ineq3}) is satisfied as well.

\subsection{Upper and lower bounds on the probabilities}

Now, perhaps after relabelling the probabilities, suppose that lower bounds
$0\leq l_{i}\leq p_{i}$ for $i=1,\ldots,m$ as well as upper bounds $p_{i}\leq
u_{i}\leq1$ for $i=m+1,\ldots,k,$ where $1\leq m<k,$ have been provided. Again
it is required that $L_{1:m}=l_{1}+\cdots+l_{m}<1$ and we search for
conditions on the $u_{i}$ that complete the prescription of a full set of
lower bounds $l_{1},\ldots,l_{k}$ so that Theorem 1 applies. Again the
$\mathbf{l}$ and $\mathbf{u}$ vectors must satisfy (\ref{mateq1}). Let
$\mathbf{x}_{r:s}$ denote the subvector of $\mathbf{x}$ given by its
consecutive $r$-th through $s$-th coordinates and $X_{r:s}$ the sum of these
coordinates provided $r\leq s$ and be null otherwise. The following equations
hold
\begin{align*}
\mathbf{u}_{1:m}  &  \mathbf{=}\mathbf{1}_{m}+\boldsymbol{l}_{1:m}%
-L_{1:m}\mathbf{1}_{m}-L_{m+1:k}\mathbf{1}_{m}\\
\mathbf{u}_{m+1:k}  &  \mathbf{=}\mathbf{1}_{k-m}-L_{1:m}\mathbf{1}%
_{k-m}+(I_{k-m}-\mathbf{1}_{k-m}\mathbf{1}_{k-m}^{\prime})\boldsymbol{l}%
_{m+1:k}.
\end{align*}
Rearranging these equations so the knowns are on the left and the unknowns are
on the right gives%
\begin{align}
\boldsymbol{l}_{1:m}+(1-L_{1:m})\mathbf{1}_{m}  &  =\mathbf{u}_{1:m}%
+L_{m+1:k}\mathbf{1}_{m}\label{mateq2}\\
\mathbf{u}_{m+1:k}-(1-L_{1:m})\mathbf{1}_{k-m}  &  =(I_{k-m}-\mathbf{1}%
_{k-m}\mathbf{1}_{k-m}^{\prime})\boldsymbol{l}_{m+1:k}. \label{mateq3}%
\end{align}
It follows from (\ref{mateq3}) that%
\begin{align}
&  \boldsymbol{l}_{m+1:k}=(I_{k-m}-\mathbf{1}_{k-m}\mathbf{1}_{k-m}^{\prime
})^{-1}[\mathbf{u}_{m+1:k}-(1-L_{1:m})\boldsymbol{l}_{k-m}]\nonumber\\
&  =(I_{k-m}-(k-m-1)^{-1}\mathbf{1}_{k-m}\mathbf{1}_{k-m}^{\prime}%
)[\mathbf{u}_{m+1:k}-(1-L_{1:m})\boldsymbol{l}_{k-m}] \label{mateq4}%
\end{align}
and substituting this into (\ref{mateq2}) gives the solution for
$\mathbf{u}_{1:m}$ as well.

So it is only necessary to determine what additional conditions have to be
imposed on the $l_{1},\ldots,l_{m},u_{m},\ldots,u_{k}$ so that Theorem 1
applies. Note that it follows from (\ref{mateq2}) that $\mathbf{u}_{1:m}$
takes the correct form, as given by (\ref{upperbd}), so it is really only
necessary to check that $\boldsymbol{l}$ is appropriate.

First it is noted that it is necessary that $k-m>1.$ The case $k-m=1$ only
occurs when $m=k-1$ and then $p_{k}=1-p_{1}-\cdots-p_{k-1}\leq1-l_{1}%
-\cdots-l_{k-1}$ which is the required value for $u_{k}$ for Theorem 1 to
apply. So when $k-m=1$ there is no choice but to put $u_{k}=1-l_{1}%
-\cdots-l_{k-1}$ and choose a lower bound for $p_{k},$ which of course could
be 0, which means that Theorem 1 applies. It is assumed hereafter that
$k-m>1.$

Now $L_{1:k}=L_{1:m}+L_{m+1:k}$ and the requirement $0\leq L_{1:k}<1$ imposes
the requirement $0\leq L_{m+1:k}<1-L_{1:m}.$ Using (\ref{mateq4}) gives%
\begin{align*}
L_{m+1:k}  &  =\mathbf{1}_{k-m}^{\prime}\boldsymbol{l}_{m+1:k}=\left(
1-\frac{k-m}{k-m-1}\right)  (U_{m+1:k}-(k-m)(1-L_{1:m}))\\
&  =\frac{(k-m)(1-L_{1:m})-U_{m+1:k}}{k-m-1}%
\end{align*}
and therefore $0\leq L_{m+1:k}<1-L_{1:m}$ iff
\begin{equation}
1-L_{1:m}<U_{m+1:k}\leq(k-m)(1-L_{1:m}). \label{ineq4}%
\end{equation}
It is seen that (\ref{ineq4}) generalizes (\ref{ineq2}) on taking $m=0.$ Now
for $i>m$%
\begin{align}
l_{i}  &  =u_{i}-(1-L_{1:m})-\frac{U_{m+1:k}}{k-m-1}+\frac{(k-m)(1-L_{1:m}%
)}{k-m-1}\nonumber\\
&  =u_{i}+\frac{(1-L_{1:m})-U_{m+1:k}}{k-m-1} \label{eq3}%
\end{align}
and so, for $i=m+1,\ldots,k,$ this implies that $l_{i}\geq0$ iff
\begin{equation}
u_{i}\geq\frac{U_{m+1:k}-(1-L_{1:m})}{k-m-1}. \label{ineq5}%
\end{equation}
So (\ref{ineq5}) generalizes (\ref{ineq2}) on taking $m=0.$ Also, if
(\ref{ineq4}) is satisfied, then $l_{i}\leq u_{i}$ for $i=m+1,\ldots,k.$

The above argument establishes the following result.\smallskip

\noindent\textbf{Theorem 3.} For $m$ satisfying $1\leq m\leq k-2,$ specifying
the bounds \newline(i) $l_{i}\leq p_{i}$ with $l_{i}\in\lbrack0,1]$ for
$i=1,\ldots,m,$ satisfying $L_{1:m}<1$ and \newline(ii) $u_{i}\geq p_{i}$ with
$u_{i}\in\lbrack0,1]$ for $i=m+1,\ldots,k,$ satisfying (\ref{ineq4}) and
(\ref{ineq5}), \newline determines\ the lower bounds $l_{m+1},\ldots,l_{k},$
given by (\ref{eq3}), which, together with $l_{1},\ldots,l_{m},$ determine the
simplex $S(\mathbf{a}_{1},\ldots,\mathbf{a}_{k})$ defined in Theorem 1.

\subsection{Determining the Elicited Prior}

So now suppose there is an elicited set of bounds that lead to the simplex
specified by Theorem 1 and it is necessary to determine the Dirichlet$(\alpha
_{1},\ldots,\alpha_{k})$ prior, denoted $\Pi_{(\alpha_{1},\ldots,\alpha_{k}%
)},$ such that $\Pi_{(\alpha_{1},\ldots,\alpha_{k})}(S(\mathbf{a}_{1}%
,\ldots,\mathbf{a}_{k}))=\gamma.$ Again we pick a point $\xi=(\xi_{1}%
,\ldots,\xi_{k})\in S(\mathbf{a}_{1},\ldots,\mathbf{a}_{k})$ and place the
mode at $\xi,$ so $\xi_{i}=(\alpha_{i}-1)/\tau$ for $i=1,\ldots,k$ with
$\tau=\alpha_{1}+\cdots+\alpha_{k}-k.$ For example, $\xi=CS(\mathbf{a}%
_{1},\ldots,\mathbf{a}_{k})$ would often seem like a sensible choice and then
only $\tau$ needs to be determined. There is a 1-1 correspondence between
$(\alpha_{1},\ldots,\alpha_{k})$ and $(\xi_{1},\ldots,\xi_{k},\tau)$ given by
$\alpha_{i}=1+\tau\xi_{i}.$

Again it makes sense to proceed via an iterative algorithm to determine $\tau
$. Provided $\Pi_{(1,\ldots,1)}(S(\mathbf{a}_{1},\ldots,\mathbf{a}_{k}%
))\leq\gamma,$ set $\tau_{0}=0$ and find $\tau_{1}$ such that $\Pi
_{(1+\tau_{i}\xi_{1},\ldots,1+\tau_{i}\xi_{k})}(S(\mathbf{a}_{1}%
,\ldots,\mathbf{a}_{k}))\geq\gamma.$ As before set $\tau_{2}=(\tau_{1}%
+\tau_{0})/2$ and then the algorithm proceeds via bisection. Determining
$\Pi_{(1+\tau_{i}\xi_{1},\ldots,1+\tau_{i}\xi_{k})}(S(\mathbf{a}_{1}%
,\ldots,\mathbf{a}_{k}))$ at each step becomes problematical even for $k=3$.
In the approach adopted here this probability content was estimated via a
Monte Carlo sample from the relevant Dirichlet. This is seen to work quite
well as, in the case of determining a prior, high accuracy for the
computations is not required.

Consider an example.\smallskip

\noindent\textbf{Example 3}. \textit{Determining a Dirichlet}$(\alpha
_{1},\alpha_{2},\alpha_{3},\alpha_{4})$\textit{ prior.}

Suppose that $k=4$ and the lower bounds $l_{1}=0.2,l_{2}=0.2,l_{3}%
=0.3,l_{4}=0.2$ are placed on the probabilities. This results in the bounds
$0.2\leq p_{1}\leq0.3,0.2\leq p_{2}\leq0.3,0.3\leq p_{3}\leq0.4,$ and $0.2\leq
p_{4}\leq0.3$ which are reasonably tight. The mode was placed at the centroid
$\xi=(0.22,0.22,0.32,0.22).$ For $\gamma=0.99,$ an error tolerance of
$\epsilon=0.005$ and a Monte Carlo sample of size of $N=10^{3}$ at each step,
the values $\tau=2560$ and $(\alpha_{1},\alpha_{2},\alpha_{3},\alpha
_{4})=(577.0,577.0,833.0,577.0)$ were obtained after 13 iterations. The prior
content of $S(\mathbf{a}_{1},\mathbf{a}_{2},\mathbf{a}_{3},,\mathbf{a}_{4})$
was estimated to be $0.989$. If greater accuracy is required then $N$ can be
increased and/or $\epsilon$ decreased.

This choice of lower bounds results in a fairly concentrated prior as is
reflected in the plots of the marginals in Figure 1. This is reflected also in
Figure 2 where scatter plots are provided of a sample of 300 from the joint
distribution for the pairs of probabilities $(p_{1},p_{2}),(p_{2},p_{3})$ and
$(p_{3},p_{4})$. This concentration is not a defect of the elicitation as
(\ref{upperbd}) indicates that it must occur when the sum of the bounds is
close to 1. So the concentration is forced by the dependencies among the
probabilities.%
%TCIMACRO{\FRAME{ftbpFU}{2.9871in}{2.7242in}{0pt}{\Qcb{Plots of the marginal
%densities determined when specifying the lower bounds $l_{1}=0.2,l_{2}%
%=0.2,l_{3}=0.3,l_{4}=0.2$ in Example 3.}}{}{figure1.eps}%
%width 2.9871in;  height 2.7242in;  depth 0pt;  original-width 6.8069in;
%original-height 6.2033in;  cropleft "0";  croptop "1";  cropright "1";
%cropbottom "0";  filename 'figure1.eps';file-properties "XNPEU";}}}%
%BeginExpansion
\begin{figure}
[ptb]
\begin{center}
\includegraphics[
height=2.7242in,
width=2.9871in
]%
{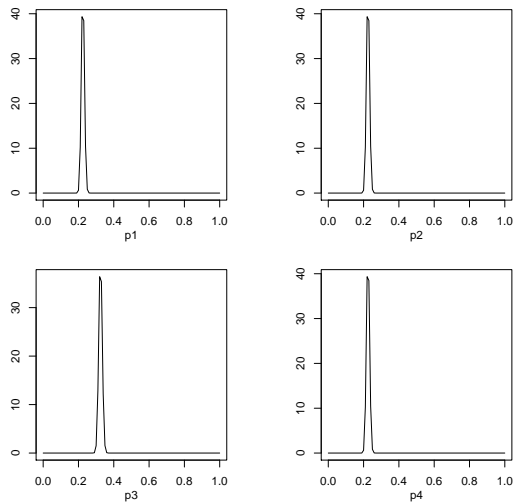}%
\caption{Plots of the marginal densities determined when specifying the lower
bounds $l_{1}=0.2,l_{2}=0.2,l_{3}=0.3,l_{4}=0.2$ in Example 3.}%
\end{center}
\end{figure}
%EndExpansion%
%TCIMACRO{\FRAME{ftbpFU}{3.8346in}{1.8645in}{0pt}{\Qcb{Scatterplots of a sample
%of size 200 from the prior in Example 2 for the pairs $(p_{1},p_{2}%
%),(p_{1},p_{3})$ and $(p_{1},p_{4})$. }}{}{figure2.eps}%
%width 3.8346in;  height 1.8645in;  depth 0pt;  original-width 8.003in;
%original-height 3.8778in;  cropleft "0";  croptop "1";  cropright "1";
%cropbottom "0";  filename 'figure2.eps';file-properties "XNPEU";}}}%
%BeginExpansion
\begin{figure}
[ptb]
\begin{center}
\includegraphics[
height=1.8645in,
width=3.8346in
]%
{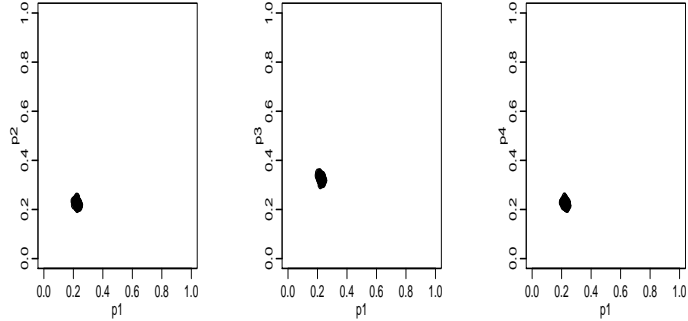}%
\caption{Scatterplots of a sample of size 200 from the prior in Example 3 for
the pairs $(p_{1},p_{2}),(p_{1},p_{3})$ and $(p_{1},p_{4})$. }%
\end{center}
\end{figure}
%EndExpansion

Consider now another example.\smallskip\newpage

\noindent\textbf{Example 4}. \textit{Determining a Dirichlet}$(\alpha
_{1},\alpha_{2},\alpha_{3},\alpha_{4},\alpha_{5},\alpha_{6},\alpha_{7}%
,\alpha_{8},\alpha_{9})$\textit{ prior.}

Suppose that $k=9$ and the lower bounds $l_{1}=0.02,l_{2}=0.02,l_{3}%
=0.0,l_{4}=0.00,l_{5}=0.00,l_{6}=0.00,l_{7}=0.10,l_{8}=0.10,,l_{9}=0.00$ are
placed on the probabilities. This leads to the following bounds for the
probabilities.%
\[%
\begin{tabular}
[c]{ccc}%
$0.02\leq p_{1}\leq0.78$ & $0.02\leq p_{2}\leq0.78$ & $0.00\leq p_{3}\leq
0.76$\\
$0.00\leq p_{4}\leq0.76$ & $0.00\leq p_{5}\leq0.76$ & $0.00\leq p_{6}\leq
0.76$\\
$0.10\leq p_{7}\leq0.86$ & $0.10\leq p_{8}\leq0.86$ & $0.00\leq p_{9}\leq0.76$%
\end{tabular}
\ \ \ \ \
\]
The mode was placed at the centroid $\xi
=(0.1,0.1,0.08,0.08,0.08,0.08,0.18,0.18,\newline0.08).$ For $\gamma=0.99,$ an
error tolerance of $\epsilon=0.005$ and a Monte Carlo sample of size of
$N=10^{3}$ at each step, the values $\tau=96$ and $(\alpha_{1},\alpha
_{2},\alpha_{3},\alpha_{4},\alpha_{5},\alpha_{6},\alpha_{7},$\newline%
$\alpha_{8},\alpha_{9})=(11.03,11.03,9.11,9.11,9.11,9.11,18.71,18.71,9.11)$
were obtained after 7 iterations. The prior content of $S(\mathbf{a}%
_{1},\mathbf{\ldots},\mathbf{a}_{9})$ was estimated to be $0.987.$ Figure 3 is
a plot of the 9 marginal priors for the $p_{i}.$ Again the dependencies among
the $p_{i}$ make the marginal priors quite concentrated.%
%TCIMACRO{\FRAME{ftbpFU}{3.1886in}{3.1765in}{0pt}{\Qcb{Plot of the 9 marginal
%priors in Example 4.}}{}{figure3.eps}{\special{ language "Scientific Word";
%type "GRAPHIC";  maintain-aspect-ratio TRUE;  display "USEDEF";
%valid_file "F";  width 3.1886in;  height 3.1765in;  depth 0pt;
%original-width 6.8329in;  original-height 6.8044in;  cropleft "0";
%filename 'figure3.eps';file-properties "XNPEU";}}}%
%BeginExpansion
\begin{figure}
[ptb]
\begin{center}
\includegraphics[
height=3.1765in,
width=3.1886in
]%
{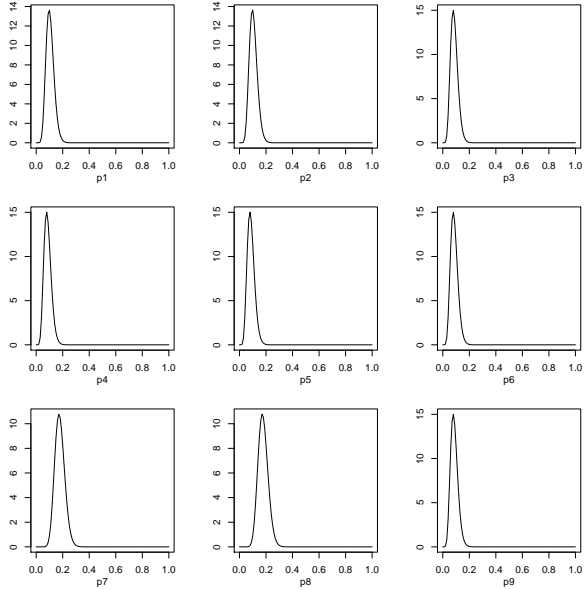}%
\caption{Plot of the 9 marginal priors in Example 4.}%
\end{center}
\end{figure}
%EndExpansion
\smallskip\smallskip

\noindent\textbf{Example 1.} \textit{(continued) Choosing the prior.}

Given that we wish to assess independence, it is necessary that any elicited
prior include independence as a possibility so this is not ruled out a priori.
A natural elicitation is to specify valid bounds (namely, bounds that satisfy
our theorems) on the $p_{i\cdot}$ and the $p_{\cdot j}$ and then use these to
obtain bounds on the $p_{ij}$ which in turn leads to the prior. So suppose
valid bounds have been specified that lead to the lower bounds $a_{i}\leq
p_{i\cdot},b_{j}\leq p_{\cdot j}.$ Then it is necessary that $l_{ij}%
=a_{i}b_{j}$ is the lower bound on $p_{ij}.$ Note that it is immediate that
the $l_{ij}$ satisfy the conditions of Theorem 1 and from (\ref{upperbd}),
$p_{ij}\leq1-\sum_{r,s}l_{rs}+l_{ij}=1-\sum_{r}a_{r}\sum_{s}b_{s}+a_{i}b_{j}$
which is greater than $l_{ij}=a_{i}b_{j}$ since $0\leq\sum_{r}a_{r}<1$ and
$0\leq\sum_{s}b_{s}<1.$ As such the region for the $p_{ij}$ contains elements
of $H_{0}.$

For this example, the lower bounds $a_{1}=0.1,a_{2}=0.0,a_{3}=0.5,b_{1}%
=0.2,b_{2}=0.2,b_{3}=0.0$ were chosen which leads to the lower bounds
\[
L=\left(
\begin{array}
[c]{ccc}%
0.02 & 0.02 & 0.00\\
0.00 & 0.00 & 0.00\\
0.10 & 0.10 & 0.00
\end{array}
\right)
\]
on the $p_{ij}.$ Note that these are precisely the bounds used in Example 4 so
the prior is as determined in that example where the indexing is row-wise.

\section{Measuring Bias in the Prior}

Here we specialize the developments discussed in Evans (2015) to the
multinomial problem with a Dirichlet prior. Suppose a quantity $\psi
=\Psi(p_{1},\ldots,p_{k})$ is of interest and there is a need to assess the
hypothesis $H_{0}:\Psi(p_{1},\ldots,p_{k})=\psi_{0}.$ Let $\pi_{\Psi}$ denote
the prior density and $\pi_{\Psi}(\cdot\,|\,f_{1},\ldots,f_{k})$ denote the
posterior density of $\Psi,$ where $(f_{1},\ldots,f_{k})$ gives the observed
cell counts. When $\Psi(p_{1},\ldots,p_{k})=(p_{1},\ldots,p_{k}),$ then
$\pi_{\Psi}$ is the Dirichlet$(\alpha_{1},\ldots,\alpha_{k})$ density and
$\pi_{\Psi}(\cdot\,|\,f_{1},\ldots,f_{k})$ is the Dirichlet$(\alpha_{1}%
+f_{1},\ldots,\alpha_{k}+f_{k})$ density. The relative belief ratio $RB_{\Psi
}(\psi_{0}\,|\,f_{1},\ldots,f_{k})$ is defined as the limiting ratio of the
posterior probability of a set containing $\psi_{0}$ to the prior probability
of this set where the limit is taken as the set converges (nicely) to the
point $\psi_{0}$. Whenever $\pi_{\Psi}(\psi_{0})>0$ and $\pi_{\Psi}$ is
continuous at $\psi_{0},$ then $RB_{\Psi}(\psi_{0}\,|\,f_{1},\ldots,f_{k}%
)=\pi_{\Psi}(\psi_{0}\,|\,f_{1},\ldots,f_{k})/\pi_{\Psi}(\psi_{0}).$ As such
$RB_{\Psi}(\psi_{0}\,|\,f_{1},\ldots,f_{k})$ is measuring how beliefs about
$\psi_{0}$ have changed from a priori to a posteriori and is a measure of
evidence concerning $H_{0}.$ If $RB_{\Psi}(\psi_{0}\,|\,f_{1},\ldots
,f_{k})>1,$ then there is evidence that $H_{0}$ is true, as belief in the
truth of $H_{0}$ has increased, if $RB_{\Psi}(\psi_{0}\,|\,f_{1},\ldots
,f_{k})<1,$ then there is evidence that $H_{0}$ is false, as belief in the
truth of $H_{0}$ has decreased and if $RB_{\Psi}(\psi_{0}\,|\,f_{1}%
,\ldots,f_{k})=1,$ then there is no evidence either way.

Given that there is a measure of evidence for $H_{0}$, it is possible to
assess the bias in the prior with respect to $H_{0}$. For this let
$M(\cdot\,|\,\psi_{0})$ denote the prior predictive distribution of
$(f_{1},\ldots,f_{k})$ given that $\Psi(p_{1},\ldots,p_{k})=\psi_{0}.$ The
bias against $H_{0}$ is assessed by%
\begin{equation}
M(RB_{\Psi}(\psi_{0}\,|\,f_{1},\ldots,f_{k})\leq1\,|\,\psi_{0}),
\label{biasag}%
\end{equation}
the prior probability that evidence in favor of $H_{0}$ will not be obtained
when $H_{0}$ is true. If (\ref{biasag}) is large, then there is bias in the
prior against $H_{0}$ and, as such, if evidence against $H_{0}$ is obtained
after seeing the data, then this should have little impact. In essence the
ingredients of the study are such that it is not meaningful to find evidence
against $H_{0}.$ To measure bias in favor of $H_{0},$ let $\psi_{\ast}$ be a
value of $\Psi$ that is just meaningfully different than $\psi_{0}.$ In other
words values $\psi$ that differ from $\psi_{0}$ less than $\psi_{\ast}$ does,
are not considered as practically different than $\psi_{0}.$ Then the bias in
favor of $H_{0}$ is measured by%
\begin{equation}
M(RB_{\Psi}(\psi_{0}\,|\,f_{1},\ldots,f_{k})\geq1\,|\,\psi_{\ast}).
\label{biasfor}%
\end{equation}
If (\ref{biasfor}) is large, then there is bias in favor of $H_{0}\ $and if
evidence in favor of $H_{0}.$is obtained after seeing the data, then this
should have little impact. It is shown in Evans (2015) that both
(\ref{biasag}) and (\ref{biasfor}) converge to 0 as $n\rightarrow\infty.$ So
bias can be controlled by sample size.

The computation of (\ref{biasag}) and (\ref{biasfor}) can be difficult in
certain contexts with the primary issue being the need to generate from the
conditional prior predictives of the data. As in the following example,
however, great accuracy is typically not required for these computations and
so effective methods are available.\smallskip

\noindent\textbf{Example 1.} \textit{(continued) Measuring bias and choosing
}$\delta$\textit{.}

To assess independence between $X$ and $Y,$ the marginal parameter
\begin{equation}
\psi=\Psi(p_{11},p_{12},\ldots,p_{kl})=\sum_{i,j}p_{ij}\ln(p_{ij}/p_{i\cdot
}p_{\cdot j}) \label{kl}%
\end{equation}
is used. Note that (\ref{kl}) is the minimum Kullback-Leibler distance between
the $p_{ij}$ values and an element of $H_{0}.$ Furthermore, $\psi=0$ iff
independence holds.

As discussed previously, it is necessary to specify a $\delta>0$ such that a
practically meaningful lack of independence occurs iff the true value
$\psi\geq\delta.$ One approach is to specify a $\delta$ such that, if
$-\delta\leq(p_{ij}-p_{i\cdot}p_{\cdot j})/p_{ij}<\delta$ for all $i$ and $j,$
then any such deviation is practically insignificant, as the relative errors
are all bounded by $\delta.$ Using $\ln(1+x)\approx x$ for small $x,$ this
condition implies that $-\delta\leq\psi<\delta.$ The range of $\psi$ is then
discretized using this $\delta$ and the hypothesis to be assessed is now,
because $\psi\geq0$ always$,$ $H_{0}:0\leq\psi<\delta.$ This assessment is
carried out using the relative belief ratios based on the discretized prior
and posterior of $\Psi$ as discussed in Section 5$.$ For the data in this
problem we take $\delta=0.01$ which corresponds to a $1\%$ relative error. So
this says that we do not consider independence as failing when the true
probabilities differ from probabilities based on independence with a relative
error of less than 1\%.

With this choice of $\delta$ the issue of bias is now addressed. The prior
distribution of the discretized $\Psi$ is determined by simulation. For this,
generate the $p_{ij}$ from the elicited prior and compute $\psi$ and the prior
probability contents of the intervals for $\psi$ given by $[0,\delta
),[\delta,2\delta),\ldots,[(k-1)\delta,k\delta)$ where $k$ is determined so as
to cover the full range of observed generated values of $\psi.$ The plot of
the prior density histogram for $\psi$ is provided in Figure 4.%
%TCIMACRO{\FRAME{ftbpFU}{2.6948in}{2.6948in}{0pt}{\Qcb{Plot of the prior
%density histogram for $\psi$ in Example 1.}}{}{figure4.eps}%
%width 2.6948in;  height 2.6948in;  depth 0pt;  original-width 6.8069in;
%original-height 6.8069in;  cropleft "0";  croptop "1";  cropright "1";
%cropbottom "0";  filename 'figure4.eps';file-properties "XNPEU";}}}%
%BeginExpansion
\begin{figure}
[ptb]
\begin{center}
\includegraphics[
height=2.6948in,
width=2.6948in
]%
{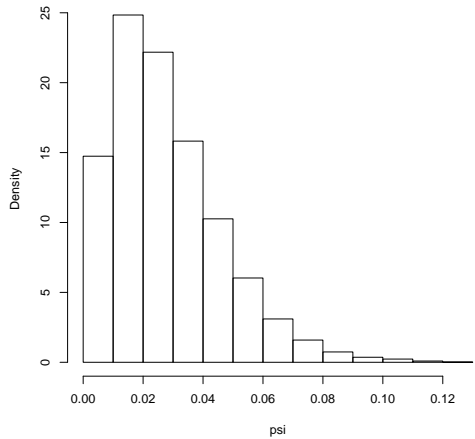}%
\caption{Plot of the prior density histogram for $\psi$ in Example 1.}%
\end{center}
\end{figure}
%EndExpansion
For inference the posterior contents of these intervals are also determined
via simulating from the posterior based on the observed data. For measuring
bias, however, we proceed as follows. Each time a generated $\psi$ satisfies
$[0,\delta)$ the corresponding $p_{ij}$ are used to generate a new data set
$F_{ij}$ and $RB_{\Psi}([0,\delta)\,|\,F_{11},\ldots,F_{kl})$ is determined
and note that this requires generating from the posterior based on the
$F_{ij}.$ The probability $M(RB_{\Psi}([0,\delta)\,|\,F_{11},\ldots
,F_{kl})\leq1\,|\,[0,\delta))$ is then estimated by the proportion of these
relative belief ratios that are less than or equal to 1. This gives an
estimate of the bias against $H_{0}.$ Estimating the bias in favor of $H_{0}$
proceeds similarly, but now the $F_{ij}$ are generated whenever $\psi
\in\lbrack\delta,2\delta)$ is satisfied, as these represent values that
correspond to just differing from independence meaningfully.

Clearly this procedure could be computationally quite demanding if highly
accurate estimates of the biases are required. In general, however, high
accuracy is not necessary. Even accuracy to one decimal place will provide a
clear indication of whether or not there is serious bias. In this problem the
biases for the elicited prior are estimated to be $0.12$ for bias for and
$0.02$ for bias against. So while there is some bias in favor of $H_{0},$ it
is not serious and there is virtually no bias against $H_{0}.$ These values
depend on the chosen value of $\delta$ but in fact are reasonably robust to
this choice. The prior probability content of the interval $[0,0.01)$ is
$0.14$ while $[0.01,0.02)$ contains $0.25$ of the prior probability. So there
is a reasonable amount of prior probability allocated to effective
independence and also to the smallest nonindependence of interest.

\section{Checking for Prior-Data Conflict}

Anytime a prior is used it is reasonable to question whether or not the prior
is contradicted by the data. For the elicitation could be in error, namely,
what if the true probabilities lie well outside the intervals obtained. If the
data demonstrate this in a reasonably conclusive way, then it would seem
incorrect to proceed with an analysis based on this prior unless there was an
absolute conviction that the amount of data was sufficient to overwhelm the
influence of the prior. Such a situation is referred to as a prior-data
conflict and methods exist to check whether or not this exists as well as
methods to deal with it.

To check for prior-data conflict we follow Evans and Moshonov (2006) and
compute the tail probability%
\begin{equation}
M(m(F_{1},\ldots,F_{k})\leq m(f_{1},\ldots,f_{k})) \label{pchk1}%
\end{equation}
where $(f_{1},\ldots,f_{k})$ is the observed value of the minimal sufficient
statistic and $M$ is the prior predictive distribution of this statistic with
density $m.$ Evans and Jang (2011a) prove that quite generally (\ref{pchk1})
converges to $\Pi(\pi(p_{1},\ldots,p_{k})\leq\pi(p_{1,true},\ldots
,p_{k,true}))$ as $n\rightarrow\infty,$ where $\Pi$ is the prior on
$(p_{1},\ldots,p_{k}).$ So (\ref{pchk1}) is indeed a valid check on the prior. \ 

When the prior is given by the uniform, then a simple computation shows that
(\ref{pchk1}) is equal to 1 and so there is no prior-data conflict.
Intuitively, the closer $\tau$ is to $0$, then the less information the prior
is putting into the analysis. This idea can be made precise in terms of the
weak informativity of one prior with respect to another as developed in Evans
and Jang (2011b). As such, if prior-data conflict is obtained with the prior
specified by a value of $(\xi_{1},\ldots,\xi_{k},\tau),$ then this prior can
be replaced by a prior that is weakly informative with respect to it so that
the conflict can be avoided and this entails choosing a value $\tau^{\prime
}<\tau.$\smallskip

\noindent\textbf{Example 1.} \textit{(continued) Checking the elicited prior.}

For the elicited Dirichlet prior the value of (\ref{pchk1}) is approximately
equal to 1 (to the accuracy of the computations) and so there is definitely no
prior-data conflict.

\section{Inference}

For data $(f_{1},\ldots,f_{k})$ and Dirichlet$(\alpha_{1},\ldots,\alpha_{k})$
prior the posterior, of $(p_{1},\ldots,p_{k})$ is Dirichlet$(\alpha_{1}%
+f_{1},\ldots,\alpha_{k}+f_{k}).$ As such it is easy to generate from the
posterior of $\psi$, estimate the posterior contents of the intervals
$[(i-1)\delta,i\delta)$ and then estimate the relative belief ratios
$RB_{\Psi}([(i-1)\delta,i\delta)\,|\,f_{1},\ldots,f_{k}).$ From this a
relative belief estimate of the discretized $\psi$ can be obtained and various
hypotheses assessed for this quantity.

As discussed in Evans (2015) the strength of the evidence provided by
$RB_{\Psi}(\psi_{0}\,|\,f_{1},\ldots,f_{k})$ is measured by
\begin{equation}
\Pi_{\Psi}(RB_{\Psi}(\psi\,|\,f_{1},\ldots,f_{k})\leq RB_{\Psi}(\psi
_{0}\,|\,f_{1},\ldots,f_{k})\,|\,f_{1},\ldots,f_{k}), \label{strength}%
\end{equation}
namely, the posterior probability that the true value of $\psi$ has a relative
belief ratio no greater than the hypothesized value. When $RB_{\Psi}(\psi
_{0}\,|\,f_{1},\ldots,f_{k})<1,$ so there is evidence against $\psi_{0},$ a
small value for (\ref{strength}) implies there is strong evidence against
$\psi_{0}$ since there is a large posterior probability that the true value
has a larger relative belief ratio than $\psi_{0}.$ When $RB_{\Psi}(\psi
_{0}\,|\,f_{1},\ldots,f_{k})>1,$ so there is evidence in favor of $\psi_{0},$
a large value for (\ref{strength}) indicates there is strong evidence in favor
of $\psi_{0}$ since there is a small posterior probability that the true value
has a larger relative belief ratio than $\psi_{0}.$ Note that when $RB_{\Psi
}(\psi_{0}\,|\,f_{1},\ldots,f_{k})>1,$ then the best estimate of $\psi$ in the
set $\{\psi:RB_{\Psi}(\psi\,|\,f_{1},\ldots,f_{k})\leq RB_{\Psi}(\psi
_{0}\,|\,f_{1},\ldots,f_{k})\}$ is $\psi_{0}$ as it has the most evidence in
its favor. Note that while the measure of strength looks like a $p$-value, it
has a very different interpretation and it is not measuring evidence.

Given that there is no prior-data conflict with the elicited prior and little
or no bias in this prior relative to the hypothesis $H_{0}$ of independence,
we can proceed to inference. \smallskip

\noindent\textbf{Example 1.} \textit{(continued) Inference.}

The posterior of the $p_{ij}$ is the
Dirichlet$(998.2,694.2,146.48,395.48,428.48,$\newline%
$96.48,2918.1,2651.1,582.48)$ distribution. For the hypothesis $H_{0}$ of
independence between the variables, and using the discretized Kullback-Leibler
divergence with $\delta=0.01,$ the value $RB_{\Psi}([0,\delta)\,|\,f_{1}%
,\ldots,f_{k})=7.13$ was obtained so there is evidence in favor of $H_{0}.$
For the strength of this evidence the value of (\ref{strength}) equals $1.$ So
the evidence in favor of $H_{0}$ is of the maximum possible strength. Of
course, this is due to the large sample size and the fact that the posterior
distribution concentrates entirely in $[0,\delta).$ Note that is a very
different conclusion than that obtained by the $p$-value based on the
chi-squared test.

\section{Conclusions}

A very natural and easy to use method has been developed for eliciting
Dirichlet priors based upon placing single bounds on the individual
probabilities that takes into account the dependencies among the
probabilities. Of course, there may be more information available, such as
upper and lower bounds on many of the probabilities. The price paid for this,
however, is a much more complicated region where the bulk of the prior mass is
located and even difficulties in determining what that region is. So indeed
further research into the development of elicitation algorithms for this
family of priors is warranted.

The application of this prior to an inference problem has also been
illustrated using a measure of statistical evidence, the relative belief
ratio, as a basis for the inferences. Given that a measure of evidence has
been identified, it is possible to assess the bias in the prior before
proceeding to inference. Also, the prior has been checked to see if it is
contradicted by the data. Finally, it is seen that the assessment of a
hypothesis can be different than that obtained by a standard $p$-value and, in
particular, provide evidence in favor of a hypothesis. Of course, this is
based on a well-known defect in $p$-values, namely, with a large enough sample
a failure of the hypothesis of no practical importance can be detected. The
solution to this problem is to say what difference matters and use an approach
that incorporates this. Relative belief inferences are seen to do this in a
very natural way. The choice of $\delta$ is not arbitrary but is rather a
fundamental characteristic of the application. When such a $\delta$ can't be
determined it is not a failure of the inference methodology, but rather
reflects a failure of the analyst to understand an aspect of the application
that is necessary for a more refined analysis to take place.

\section{References}

\noindent Chaloner, K. and Duncan, G.T. (1987). Some properties of the
Dirichlet multinomial distribution and its use in prior elicitation.
Communications in Statistics -- Theory and Methods, 16, 511--523.\smallskip

\noindent Dickey, J. M., Jiang, J. M., and Kadane, J. B. (1987). Bayesian
methods for censored categorical data. Journal of the American Statistical
Association, 82, 773--781.\smallskip\ 

\noindent Dorp, J, and Mazzuchi, T. A. (2003) Parameter specification of the
beta distribution and its Dirichlet extensions utilizing quantiles. Handbook
of Beta Distributions and Its Applications, eds. Gupta, A. K. and Nadarajah,
3-32, S. Marcel Dekker Inc.\smallskip

\noindent Evans, M. and Moshonov, H. (2006) Checking for prior-data conflict.
Bayesian Analysis, 1, 4, 893-914.\smallskip

\noindent Evans, M. (2015) Measuring Statistical Evidence Using Relative
Belief. Monographs on Statistics and Applied Probability 144, CRC
Press.\smallskip

\noindent Evans, M. and Jang, G-H. (2011a) A limit result for the prior
predictive applied to checking for prior-data conflict. Statistics and
Probability Letters, 81, 1034-1038.\smallskip

\noindent Evans, M. and Jang, G-H. (2011b). Weak informativity and the
information in one prior relative to another. Statistical Science, 26, 3,
423-439.\smallskip\ 

\noindent Garthwaite, P. H., Kadane, J. B., and O'Hagan, A. (2005) Statistical
methods for eliciting probability distributions. Journal of the American
Statistical Association, 100, 470, 680-700.\smallskip

\noindent O'Hagan, A., Buck C. E., Daneshkhah, A., Eiser, J. R., Garthwaite,
P. H., Jenkinson, D. J., Oakley, J. E., Rakow, T. (2006) Uncertain Judgements:
Eliciting Experts' Probabilities. John Wiley \& Sons.\smallskip

\noindent Regazzini, E. and Sazonov, V.V. (1999). Approximation of laws of
multinomial parameters by mixtures of Dirichlet distributions with
applications to Bayesian inference. Acta Applicandae Mathematicae, 58,
247--264.\smallskip

\noindent Snedecor, G. and Cochran, W. (1967) \textit{Statistical Methods,
}6th ed.\textit{, }Iowa State University Press.
\end{document}